# On the relation of the sizes of trans-neptunian dwarf planets Pluto and Eris

Yury I. Rogozin

VEDA LLC, P.O. Box 98, Moscow, 125310, Russia

E- mail: yrogozin@gmail.com

**ABSTRACT**

The discovery of the largest trans-neptunian object 2003 UB313 (dwarf planet Eris) was made more than 5 years ago, but the question on the true relation of the sizes of Pluto and Eris (and according to of their densities) remains debatable in view of a sizable scatter of their size's estimates obtained by the various methods. Here, we first used a semi-empirical approach to deduce the expression linking the orbital parameter eccentricity to the physical properties of the trans-neptunian dwarf planets and have applied it to determining the mean size of these planets. In doing so is proved that the mean Eris' size should be about 9 % larger than of Pluto's. Based on the published photometric data and the derived mean diameter the possible estimates of the minimum and maximum diameters of Pluto and Eris on the assumption of a deviation their form from spherical are provided. The probable reason for an occurrence of such an aspherical form of these dwarf planets is briefly discussed.

## 1. INTRODUCTION

From November 2010 the widely known question about "Pluto or Eris is the biggest dwarf planet?" up to now remains to be solved. The reason for an appearance this question is preliminary results of the observations of a stellar occultation by Eris on November 6, 2010. It is suggests that its diameter may be only 2320 km, i.e. about of Pluto's (Brown 2010). The justified solving of this question has not only a principal scientific interest but also a large significance for a possible remedying in the future an uncertainty in the physical properties of other trans-neptunian dwarf planets (TNDPs).

It is well known that Pluto and Eris have the different history of their observations. Pluto was discovered in 1930, whereas Eris only about 6 years ago (Brown et al 2005). The most known estimate of the Pluto diameter is 2306 ± 20 km (Buie et al 2006), however there presents a diversity of other estimates. Though a study of the most distant known large object of the Solar System (97 AU) Eris represents a serious problem, but already shortly after its discovery the results of the direct *Hubble Space Telescope* (*HST*) observations were published, resulting of its diameter 2400 ± 100 km (Brown et al 2006). In the same year value 3000 ± 400 km has been obtained from the indirect radiometric observations (Bertoldi et al 2006). A consequence of this discrepancy of the Eris size became the almost double discrepancy in the estimates of its density, namely 2.3 ± 0.3 g cm$^{-3}$ in the first case and 1.2 ± 0.6 g cm$^{-3}$ in the second, following from calculated its mass 1.66 x 10$^{22}$ ± 0.02 x 10$^{22}$ kg (Brown & Schaller 2007). Result from *Nature* (Bertoldi et al 2006) corrected for the originally adopted value of ratio between the bolometric albedo (representing the total reflected energy and used in the radiometric technique) and the optical "geometric" albedo (representing the reflection in some visual wavelength and used in *HST*- observations) from 0.9 to 0.7 decreases by 100 km, "so that both measurements would agree within the 68% confidence limits at ca. 2500 km" (Bertoldi 2007). The later observations of the Kuiper belt objects with the *Spitzer Space Telescope (SST)* (Stansberry et al 2008) suggested the Eris diameter is 2600 +400/-200 km. This ambiguous situation was aggravated by the recent measurements the length of a star occultation by Eris that according to author's assessment (Sicardi et al 2011) showed of "our observation is consistent with a spherical Eris with radius $R_E$ = 1163 ± 6 km, density $\rho$=2.52 ± 0.05 g cm$^{-3}$, and visible geometric albedo $p_v$ = 0.96 +0.09/-0.04". Respectively, in view of some uncertainties of measurements of the sizes of Pluto and Eris this estimate would make Eris' size about the same that Pluto's. In doing so the validity of the foregoing previously results for the Eris' size is open to question because any published their interpretation in the light of these recent observations is lacking. On this basis, it is evident that now by solely existing methods it is difficult to make any justified inferences about the true relation of the sizes of Pluto and Eris. It seems likely that the essential drawback of all these studies is the assumption that Pluto and Eris have the spherical form. Thus, as way out some auxiliary approach is called for.

The purpose of this Letter is the analytical study of a possible influence of an aspherical form of Pluto and Eris on their measured sizes and the evaluation of the true relation of their sizes and density in view of this form.

## 2. DWARF PLANETS ECCENTRICITY TREATMENT

It is well known that a distinctive feature of the orbital parameters of TNDPs compared to classical planets is their large eccentricity *e* about 0.249 for Pluto and 0.442 for Eris. This suggests that to seek the linkage between this orbital parameter of TNDPs and their physical properties and distance from the Sun. We proceed from that at formation of the planets from a protoplanetary disk can't be any physical reasons of promoting to an occurrence of highly elliptic orbits of the planets. The large eccentricity of such orbits supposedly is a consequence of the forced transfer of the planets from one orbit to another. Therefore, on our sight, can be theorize that all planets of the Solar system with $e \approx 0.1 \div 0.5$ are the former satellites of their hosts, forced to escape their former orbits. In this connection, it is important to note that *e* of Neptune's orbit approximately 5 times less than *e* of other giant planets that can testify to its special way during the formation of the Solar system. Probably, in view of this and other Neptune's peculiarities, the planetary science still should answer to a question: "where is Neptune's dynamical birthplace, and how did it migrate to its current location?" (Agnor et al 2009). In this light, on our sight, quite probable our working hypothesis looks for Neptune is the newcomer in the Solar system, and known now TNDPs in the past (before the arrival to the Solar system) were its satellites.

Following presents the semi-empirical deduction of eccentricity equation of TNDPs versus their physical and orbital parameters. Obviously that at transfer of a planet from one constant orbit to another an eccentricity of its final orbit is formed basically under the action on a planet of two the basic forces: inertial force $F_i$ and drag force $F_d$. As a result of inertia $F_i$, *e* should to increase as mass *M* of a celestial body. By contrast, drag force $F_d$ interfere with a growth *e*. The drag force theory holds that it varies with the square of speed $v^2$ and the square of radius of a body $r^2$ (for a body of the spherical form). But from 3-rd the Kepler's law $v^2$ is proportional $1/R$, where *R* is semi-major axis of an orbit. In result, the eccentricity would be expressible as

$$e \sim F_i/F_d \sim M_n/(v_n^2 \, r_n^2) \sim M_n R_n/r_n^2 \qquad (1)$$

where symbol *n* refers to a planet of interest.

In terms of the stated considerations of Neptune and an existence of orbital resonances of TNDPs with Neptune as the nearest giant planet only way to receive a dimensionless value $e$ is symmetric introduction in Eq. (1) appropriate tabulated Neptune's parameters $M_N = 102.4 \times 10^{24}$ kg, $R_N = 30.06$ AU and $r_N = 24622$ km (http://ssd.jpl.nasa.gov/?planet_phys_par). Then we obtain full-blown the eccentricity expression as

$$e = C \frac{r_N^2}{r_n^2} \frac{R_n}{R_N} \frac{M_n}{M_N} \qquad (2)$$

where *C* is a numerical factor that based on preliminary estimates by trial-and-error method we have accepted as $\pi$.

## 3. RESULTS

First of all we have applied Eq. (2) to Pluto as to the most researched dwarf planet. The substitution in it the foregoing values of mean radius $r_n = 1153$ km, semi-major axis $R_n = 39.48$ AU and the appropriate mass $M_n = 13.034 \times 10^{21}$ kg (at $\rho = 2.03$ g cm$^{-3}$) gives $e = 0.240$. It is possible to explain this difference relative to the tabulated value $e = 0.2488$ by binary character of the system Pluto - Charon. The search of Pluto's radius satisfying to Eq. (2) at the Pluto's density 2.03 g cm$^{-3}$ (that is allowable in view of the Pluto is about 8 times more massive than the Charon) gives $r_n = 1198$ km. At this effective radius and $\rho = 2.03$ g cm$^{-3}$ the total mass of this binary system is $14.62 \times 10^{21}$ kg. On this basis the Charon's mass follows as the difference of the overall mass of system and the foregoing Pluto's mass that is $1.586 \times 10^{21}$ kg. In result given the mean Charon's radius $r_n = 603.5$ km (Sicardi et al 2006) we find that the Charon's density is 1.72 g cm$^{-3}$ that well agrees to recent value $1.71 \pm 0.08$ g cm$^{-3}$ (Sicardi et al 2006). By this means we have shown the feasibility of a use of the Eq. (2) for a finding the sizes of other TNDPs with a known mass. With use of the data about the maximum albedo of Pluto $p_v = 0.66$ its minimum radius makes up 1129 km. Now, proceeding Pluto's volume both found of mean and minimal radius, the maximal value of Pluto's radius is 1177.5 km. This value corresponds to the data of early measurements of Pluto's radius by a stellar occultation $1180 \pm 5$ km (Millis et al 1993). Thus, Pluto has the deviations from spherical form that is necessary to take into account by a comparison of its size with of Eris.

Similarly, Eq. (2) was used for a determination of the Eris' size. Using of the tabulated data for Eris ( $R_n = 67.67$ AU, $M_n = 1.66 \times 10^{22} \pm 0.02 \times 10^{22}$ kg, and $e = 0.44177$) its mean diameter measures $2509 \pm 14$ km and the density accordingly measures $2.007 \pm 0.012$ g cm$^{-3}$. In terms of the most probable maximum geometric albedo of Eris $p_v = 0.86$ (Brown et al 2006) its minimal diameter is 2400 km. Now, in terms of the mean diameter the maximal Eris' diameter makes up 2623 km. It is evident that the found extreme estimates of the Eris' diameter fits the most probable data of *HST* and *SST* observations fairly well. Furthermore, the mean Eris' diameter 2509 km is close to the foregoing corrected estimate of IRAM measurements ca. 2500 km.

As a further demonstration the validity of our approach to determination the sizes of TNDPs we have applied Eq. (2) to one more dwarf planet Haumea with known mass. With use of its orbital data $e = 0.18874$ and $R_n = 43.335$ AU (http://ssd.jpl.nasa.gov/dwarf planet_), and mass $M_n$ = 4.006 x $10^{21}$ kg ( Ragozzine & Brown 2009) we find that its mean diameter makes up 1508.6 km that agree with the commonly accepted estimate ~ 1500 km (Lykawka et al 2011).

## 4. DISCUSSION AND CONCLUSIONS

In summary, the relation between mean diameters of Eris and Pluto with regard to deviations from the sphericity respectively 2509±110 km and 2306±50 km is 1.087. Thus, is shown that the mean Eris' diameter almost 9 % is larger than such of Pluto's, though in consequence of an aspherical form of these TNDPs the upper estimate of Pluto's diameter 2360 km is close to the lower estimate of Eris' diameter 2400 km. On account of uncertainties of these measurements it is illusory can to result both in rough equality, and even to an overestimate of the size of Pluto versus Eris.

As indicated above, Pluto and Eris have the aspherical form. The dwarf planet Haumea the aspherical form has too. This feature of TNDPs is most likely to occur by an impact action of the small-sized asteroids and possibly comets. They were able to grind icy surface of TNDPs to some extent. Haumea with its very appreciable the asphericity of 1960 x 1580 x 996 $km^3$ is the most dramatic example of such asteroid's grinding of TNDPs. In this instance, by means of the transfer of own momentum asteroids and comets could highly cut the Haumea's rotation period (up to 3.9 h in comparison with substantially greater Eris' rotation period), that up today did not find any explanation. Possibly too, that the reason of the lack of surface $N_2$ and $CH_4$ ices at the other large Kuiper belt objects in particular Haumea by comparison with Pluto and Eris cited in (Stern 2010) is larger collisional grinding of their surface by asteroids and comets.

The value of Eris' density 2.007 g $cm^{-3}$ afforded by its mean diameter 2509 km and near value of Pluto's density are more consistent to similar abundances of surface $N_2$ and $CH_4$ ices on Pluto (97% and 3%) and Eris (90% and 10%) (Tegler et al 2010) than foregoing values of Eris 2.3 g $cm^{-3}$ and 2.52 g $cm^{-3}$ that differs noticeably from Pluto's density 2.03 g $cm^{-3}$. Thus, in our opinion found relation of the sizes of Pluto and Eris more than others is consistent to the relation of their masses at close values of their density. Thereof it is possible to assume their almost identical composition and uniform origin.

The above-stated analysis testifies that in 2006 Pluto was completely fairly demoted from a category of classic planets in terms of its smaller diameter than that of Eris. It is hoped that obtained the relation of the sizes of Eris and Pluto will make possible an improved idea of the physical properties of other TNDPs and candidates in TNDPs.

**REFERENCES**

Agnor, C. B., Barr, A. C., Bierhaus, B. et al., 2009. The Exploration of Neptune and Triton. 2. 1. Executive summary. The NRC 2009 Planetary Science Decadal Survey.

Bertoldi, F., Altenhoff, W., Weiss, A., Menten K. M., Thum, C., 2006. The trans-neptunian object $UB_{313}$. Nature 439, 563-564.

Bertoldi, F., 2007. Comment on recent *Hubble Space Telescope* measurement of 2003 UB313 by Brown et al. http://www. astro.uni-bonn.de~bertoldi/ub313.

Brown, M. E., Trujillo, C. A., Rabinowitz, D. L., 2005. Discovery of a planetary sized object in the scattered Kuiper belt. Astrophysical Journal Letters 635, L97-L100.

Brown, M. E., Schaller, E. L., Roe, H. G., Roe H. G., Rabinowitz, D. L., Trujillo, C. A., 2006. Direct measurement of the size of 2003 UB313 from the *Hubble Space* Telescope.

    Astrophysical Journal Letters 643, L61-L63.

Brown, M. E., Schaller, E. L., 2007. The mass of dwarf planet Eris. Science 316, 1585.

Brown, M. E., 2010. http: //www. caltech.edu/ ~ mbrown/, "Mike Brown's planets".

Buie, M. W., Grundy, W. M., Young, E. F., Young L. A., S. A. Stern, 2006. Orbits and photometry of Pluto's satellites: Charon, S/2005 P1, and S/2005 P2. Astronomical Journal 132, 290-298.

Lukawka, P. S., Horner, J., Nakamura, A. M. et al. 2011 Dynamical evolution of Haumea collisional family. EPSC Abstracts, vol. 6, EPSC-DPS2011-891.

Millis, R. L., Wasserman, L. H., Franz, O. G. et al., 1993. Pluto's radius and atmosphere – results from the entire 9 June 1988 occultation data set. Icarus 105, 282-297.

Ragozzine, D., Brown, M. E., 2009. Orbits and masses of the satellites of dwarf planet Haumea (2003 EL61). Astronomical Journal 137, 4766-4776.

Sicardi, B., Belucci, A., Gendron, E., et al., 2006. Charon's size and an upper limit on the atmosphere from a stellar occultation. Nature 439, 52- 54.

Sicardi, B., Ortiz, J. L., Assafin, M., et al., 2011. Size, density albedo, and atmosphere limit of dwarf planet Eris from a stellar occultation. EPSC Abstracts, vol. 6, EPSC-DPS2011-137-8.

Stansberry, J., Grundy, W., Brown, M., Cruikshank, D., Spencer, J. et al., 2008. Physical Properties of Kuiper Belt and Centaur Objects: Constraints from *Spitzer Space Telescope*. In: The Solar System beyond Neptune, Eds. M. A. Barucci, H. Boehnhardt et al (Tucson: Univ. Arizona Press), pp. 161 – 179.

Stern S. A., 2010. Pluto is again harbinger. Nature 468, 775-776.

Tegler, S. C., Cornelison, D. M., Grundi W. M. et al. 2010. Methane and nitrogen abundances on Pluto and Eris. Astrophysical Journal 725, 1296-1305.